\documentclass[aps,prb,twocolumn,superscriptaddress]{revtex4}
\usepackage{graphicx}
\usepackage{bm}

\begin{document}

\preprint{}

\title{
NMR evidence for an intimate relationship between antiferromagnetic spin fluctuations  and extended s-wave superconductivity in mono-crystalline SrFe$_2$(As$_{1-x}$P$_{x}$)$_2$
}

\author{M. Miyamoto}
\author{H. Mukuda}
\email[]{e-mail  address: mukuda@mp.es.osaka-u.ac.jp}
\affiliation{Graduate School of Engineering Science, Osaka University, Osaka 560-8531, Japan}

\author{T. Kobayashi}
\affiliation{Department of Physics, Graduate School of Science, Osaka University, Osaka 560-0043, Japan}

\author{M. Yashima}
\author{Y. Kitaoka}
\affiliation{Graduate School of Engineering Science, Osaka University, Osaka 560-8531, Japan}

\author{S. Miyasaka}
\author{S. Tajima}
\affiliation{Department of Physics, Graduate School of Science, Osaka University, Osaka 560-0043, Japan}

\date{\today}

\begin{abstract}
We report systematic $^{31}\!$P-NMR study on iron (Fe)-based superconductors SrFe$_2$(As$_{1-x}$P$_{x}$)$_2$ (Sr122AsP), in which a superconducting (SC) transition temperature $T_c$ at $x$=0.35 increases from $T_c$=26 K up to 33 K by annealing an as-grown mono-crystalline sample. The present NMR study has unraveled that $T_c$ reaches a highest value of 33 K at $x$=0.35 around a quantum critical point at which antiferromagnetic (AFM) order disappears. 
When noting that the SC transition disappears at $x$=0.6 where the AFM spin fluctuations (SFs) are no longer present, we remark that the onset and increase of $T_c$ are apparently associated with the emergence and enhancement of AFM-SFs, respectively. 
In the SC state, the residual density of state (RDOS) at the Fermi energy $E_{\rm F}$ in the SC state becomes much smaller for the annealed sample than for the as-grown one, suggesting that some inhomogeneity and/or imperfection for the latter increases RDOS as expected for unconventional SC state with nodal gap. These findings in Sr122AsP are consistent with the unconventional $s_\pm$-wave Cooper pairing state that is mediated by AFM-SFs. We also discuss  other key-ingredients besides the AFM-SFs to increase $T_c$ further.  
\end{abstract}


\maketitle

\section{Introduction} 

The discovery of iron (Fe)-based superconductor has triggered numerous research works on Fe-based layered compounds with various chemical composites and different blocking layers.\cite{Kamihara2008}
However, the diversity of experimental results on superconducting (SC) and normal-state properties prevents us from coherently understanding their SC characteristics and getting some insight into a promising mechanism to reach a highest SC transition temperature of $T_c$= 55 K. 
It is known that  the highest $T_c$  states in these {\it doped} compounds are characterized by the optimum height of pnictogen ($Pn$) from Fe-plane $h_{Pn}=1.35\sim1.38$\AA~\cite{Mizuguchi} and the optimum $Pn$-Fe-$Pn$ bonding angle $\alpha\sim$109.5$^\circ$ of regular tetrahedral Fe$Pn_4$.\cite{C.H.Lee} 
The highest $T_c$ in $R$FeAs(O,F)(denoted as $R$1111OF) with $R$=Nd and Sm takes place when antiferromagnetic (AFM) order is suppressed by varying the valence of divalent iron Fe$^{2+}$ of Fe-pnictogen layer, i.e. doping electrons through the substitution of monovalent fluorine F$^{1-}$ for divalent oxygen O$^{2-}$. 
The isovalent substitution of P for As in the FeAs layer keeping Fe$^{2+}$ also replaces an AFM parent compound to a superconductor even though the tetrahedral parameters are deviated from the optimal ones. 
In these compounds, it should be noted that $h_{Pn}$ is a key parameter to evolve from an AFM phase to two types of SC phases: The AFM order takes place for 1.32\AA $<h_{Pn}<$1.42\AA, the nodeless SC state for $h_{Pn}>$1.42\AA\ and the nodal SC one for $h_{Pn}<$1.32\AA.\cite{KinouchiPRB}

With respect to isovalent-substitution compounds such as $M$Fe$_2$(As$_{1-x}$P$_{x}$)$_2$ with $M$=Ba, Sr or Ca (denoted as $M$122AsP hereafter), the superconductivity with $T_c\sim$30 K in Ba122AsP\cite{Kasahara,NakaiPRB,NakaiPRL,Hashimoto2,JSKim,Wang,Yamashita,Yoshida2014,Shibauchi} and Sr122AsP \cite{Shi,Kobayashi,KobayashiPRB2013,Dulguun,Maeda,Murphy,Strehlow} takes place around a quantum critical point (QCP) at which AFM order disappears, exhibiting a nodal-gap structure. 
This is empirically understood because their $h_{Pn}$s are smaller than 1.32\AA.\cite{Hashimoto2,KinouchiPRB} 
Note that the lattice parameter along the c-axis in Sr122AsP is smaller than that in Ba122AsP due to the ion radius smaller for Sr than for Ba, giving rise to a significant deformation in the Fermi surface topologies.\cite{SuzukiARPES} 
Nevertheless the phase diagram of AFM and SC phases in Sr122AsP\cite{Shi,KobayashiPRB2013} resembles that in Ba122AsP.\cite{Kasahara} Furthermore, it was reported that the $T_c$ for SrFe$_2$(As$_{1-x}$P$_{x}$)$_2$ at $x$=0.35 increases from $T_c\sim$26 K up to 33 K by annealing the as-grown mono-crystalline sample.\cite{KobayashiPRB2013} 

Motivated by the diversity of these experimental results on the AFM and SC properties in $M$122AsP, in this paper, we report systematic $^{31}\!$P-NMR study on the AFM and SC properties in SrFe$_2$(As$_{1-x}$P$_{x}$)$_2$. 
The present study has revealed that $T_c$ becomes a maximum at $x$=0.35 around which the AFM order disappears, and that the large reduction of residual density of states (RDOS) at the Fermi energy $E_{\rm F}$ takes place for the annealed one. 
The latter result means that  $T_c$ increases from 26 K up to 33 K as a result of  the reduction of RDOS, where the defects are reduced by annealing. 
These experimental findings reveal that the unconventional $s_\pm$-wave Cooper pairing state is realized in Sr122AsP, which is mediated by the AFM-SFs.\cite{Mazin,Kuroki1} 
We also discuss other key-ingredients besides the AFM-SFs to increase $T_c$ further, in comparison with the higher $T_c$ Fe-pnictides.

\section{Experimental} 

Mono-crystalline samples of SrFe$_2$(As$_{1-x}$P$_{x}$)$_2$  were synthesized by self-flux method.\cite{Kobayashi,KobayashiPRB2013} 
We performed $^{31}\!$P-NMR measurements for $x$=0.1, 0.2, 0.28, 0.35, 0.5, 0.6, and 1.0 at an external field $B_0 \sim$ 11.95 T perpendicular to the $c$-axis using the aligned mono-crystalline samples. 
The Knight shift $^{31}\!K$ was determined with respect to a resonance field in H$_3$PO$_4$. 
Nuclear spin-lattice relaxation rate $1/T_1$ was obtained from the recovery of nuclear magnetization by fitting to a simple exponential recovery curve of $m(t) = [M_0-M(t)]/M_0 = \exp(-t/T_1)$ for $^{31}\!$P($I=1/2$) at $B_0 \sim$11.95 T. 
Here $M_0$ and $M(t)$ are the respective nuclear magnetizations of $^{31}\!$P for the thermal equilibrium condition and at a time $t$ after the saturation pulse.

\section{Results} 

\subsection{Normal-state properties in SrFe$_2$(As$_{1-x}$P$_{x}$)$_2$} 

\begin{figure}[b]
\centering
\includegraphics[width=7cm]{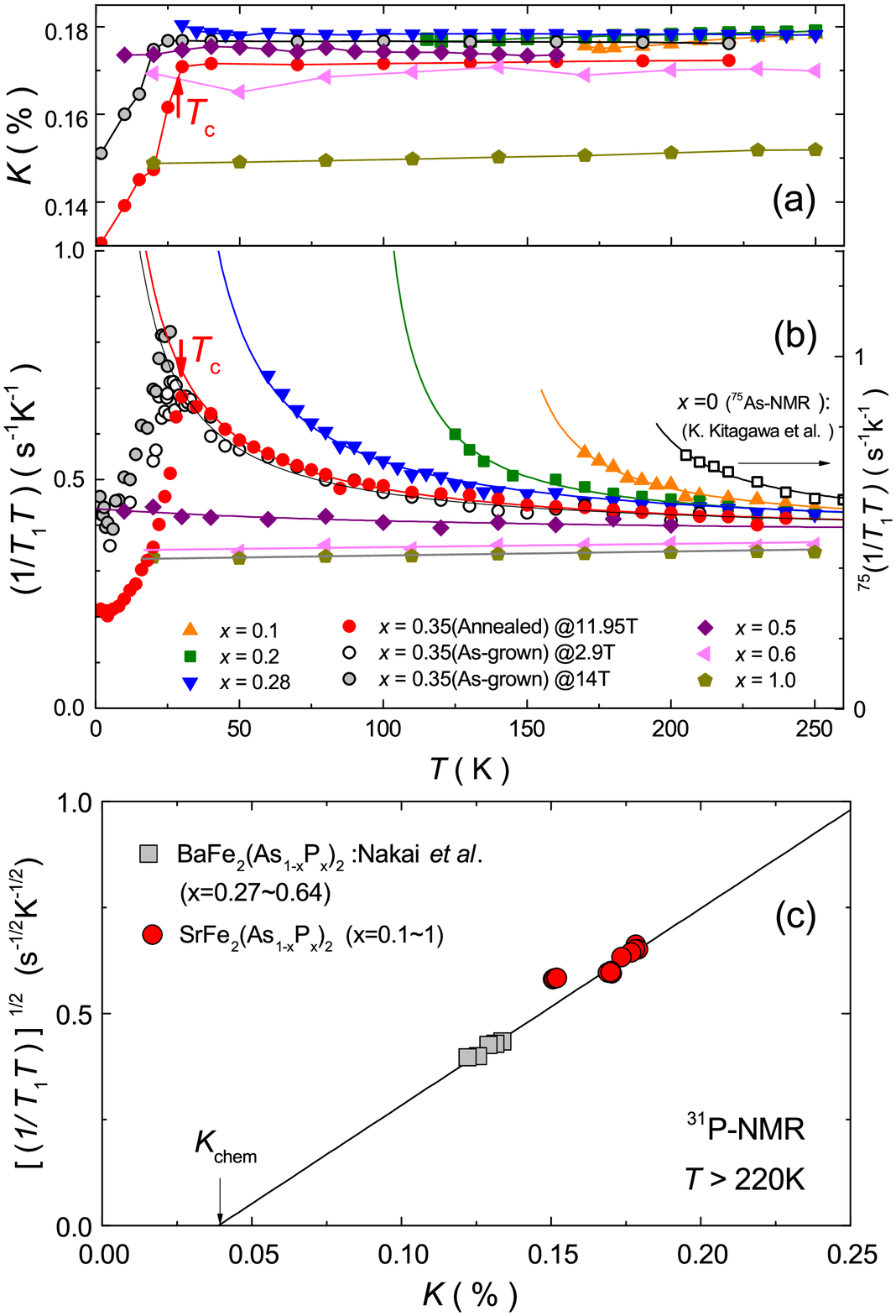}
\caption[]{(Color online) (a) $T$ dependence of Knight shift for 0.1$\le x \le$1.0 of SrFe$_2$(As$_{1-x}$P$_{x}$)$_2$. (b) $T$ dependence of $^{31}\!$P-NMR $(T_1T)^{-1}$, along with the data on $x$=0 obtained by $^{75}\!$As-NMR.\cite{KitagawaSr122}. 
(c) Plots of $\sqrt{(1/T_1T)_0}$ versus $^{31}\!K$ for $T~>~220$K. Both data for Sr122AsP and Ba122AsP\cite{NakaiPRL} are on a linear relation of $\sqrt{(1/T_1T)_{0}}=K_{\rm chem}+K_s$ with $K_{\rm chem}\sim$0.04 ($\pm$0.01)\%. Note that the $K_{\rm chem}$ of Sr122AsP coincides with those of Ba122AsP\cite{NakaiPRL} and LaFe(As$_{1-x}$P$_x$)O (La1111AsP).\cite{MukudaJPSJ2014,MukudaPRB2014} 
}
\label{1/T1TvsT}
\end{figure}

Figures \ref{1/T1TvsT} (a) and \ref{1/T1TvsT}(b) show the respective temperature ($T$) dependencies of Knight shift $K$ and $1/T_1T$ of $^{31}\!$P-NMR for Sr122AsP. 
The Knight shift $K(T)$ comprises the spin shift $K_s$ and the chemical shift $K_{\rm chem}$. 
$K_s$ is given by 
$K_s=^{31}\!$$A_{\rm hf}(q$=$0)\chi_0\propto ^{31}\!$$A_{\rm hf}(q$=$0)N(E_{\rm F})$, using the static spin susceptibility $\chi_0$ and the density of states (DOS) $N(E_{\rm F})$ at $E_{\rm F}$. 
As seen in Fig.\ref{1/T1TvsT}(a), $K(T)$s for all the samples stay constant in the normal state and a magnitude gradually decreases with increasing $x$.

By contrast, the $1/T_1T$s for $x \le$ 0.5 in Fig. \ref{1/T1TvsT}(b) develop markedly upon cooling irrespective of either the AFM or SC samples, probing the development of AFM spin fluctuations (SFs).
The $1/T_1T$ is generally described as, 
$1/T_1T\propto \sum_{\bm q} |A_{\rm hf}(\bm q)|^2 \chi''({\bm q},\omega_0)/\omega_0$,
where $A_{\rm hf}(\bm q)$ is a wave-vector ${\bm q}$-dependent hyperfine-coupling constant, $\chi({\bm q},\omega)$ a dynamical spin susceptibility, and $\omega_0$ an NMR frequency. 
Since the $1/T_1T$ in Sr122AsP  stays constant as temperature goes up in $T$-range higher than 200 K, 
we assume that $1/T_1T$ is decomposed as,
\[
1/T_1T=(1/T_1T)_{\rm AFM}+(1/T_1T)_{0}, 
\]
where the first term is a contribution relevant with AFM-SFs at a finite wave vector with either ${\bm Q_{\rm AF}}$=(0,$\pi$) or ($\pi$,0). 
This $(1/T_1T)_{\rm AFM}$ significantly develops upon cooling. $(1/T_1T)_{0}$ is a $q$-independent contribution dominated by single-particle excitations near $E_{\rm F}$. 
It is hence anticipated that $K_s\propto N(E_{\rm F})$ is proportional to $\sqrt{(1/T_1T)_{0}}$ since the Korringa's relation $(1/T_1T)\propto N(E_{\rm F})^2$ holds. 
Figure \ref{1/T1TvsT}(c) shows a plot of $\sqrt{(1/T_1T)_0}$ vs $K$ using the data at $T\ge$220 K for various samples, pointing to a linear relation of $\sqrt{(1/T_1T)_{0}}=K_{\rm chem}+K_s$ with $K_{\rm chem}\sim$0.04 ($\pm$0.01)\%. 
The $K_{\rm chem}$ of Sr122AsP coincides with those of Ba122AsP\cite{NakaiPRL} and LaFe(As$_{1-x}$P$_x$)O (La1111AsP).\cite{MukudaJPSJ2014,MukudaPRB2014}  
Here we assume that the $(1/T_1T)$  is mostly dominated by $(1/T_1T)_{0}$ at $T\ge$220 K for $x\ge$0.1.

\subsection{Evolution of electronic state and AFM-SFs in the phase diagram of SrFe$_2$(As$_{1-x}$P$_{x}$)$_2$} 

\begin{figure}[t]
\centering
\includegraphics[width=7cm]{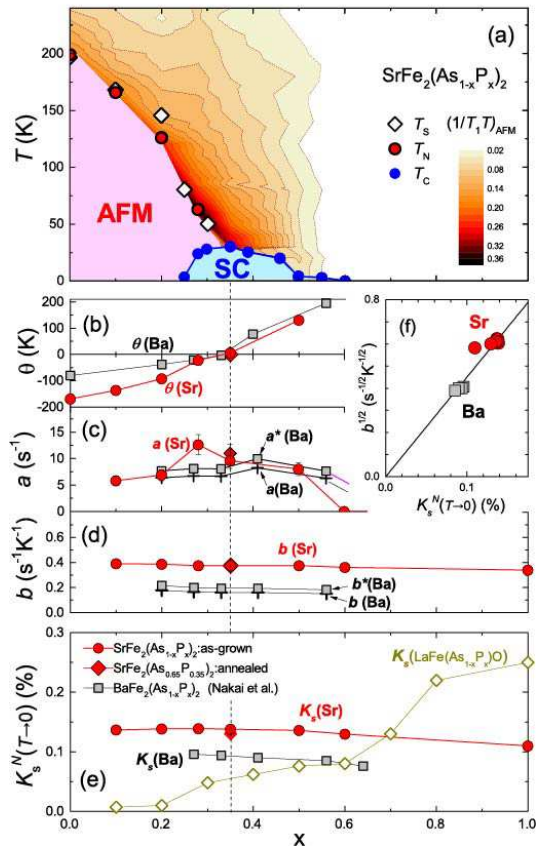}
\caption[]{(Color online) 
(a)  Phase diagram of AFM and SC states against P concentration $x$ in SrFe$_2$(As$_{1-x}$P$_x$)$_2$ along with the contour plot of $(1/T_1T)_{\rm AFM}$ in the normal state.\cite{KobayashiPRB2013} 
$T_N$ is determined by the rapid decrease of NMR intensity due to the broadening in association with the onset of AFM order. 
$T_s$ is the structural transition temperature estimated from the anomaly in resistivity.\cite{KobayashiPRB2013} 
The parameters of (b) $\theta$, (c) $a$ and (d) $b$ deduced from the simulation of $1/T_{1}T= a/(T+\theta)+b$ to the data of $1/T_1T$ in Fig. \ref{1/T1TvsT}(b).
Here, $a^*$(Ba) was evaluated from $a$(Ba)\cite{NakaiPRL} using the ratio of $(A_{\rm hf}^{\rm Sr}/A_{\rm hf}^{\rm Ba})^2$ for quantitative comparison with Sr122AsP.
(e) $x$ dependence of $K_s\propto~N(E_{\rm F})$ in the normal state along with the data of Ba122AsP\cite{NakaiPRL} and La1111AsP.\cite{MukudaJPSJ2014}
(f) Plot of $\sqrt{b}$ versus $K_s^N(T\rightarrow0)$ for Sr122AsP and Ba122AsP.\cite{NakaiPRL}
}
\label{PhaseDiagram}
\end{figure}

Figure \ref{PhaseDiagram}(a) presents the phase diagram of AFM and SC states against P concentration $x$ in SrFe$_2$(As$_{1-x}$P$_{x}$)$_2$. 
Here the contour of $(1/T_1T)_{\rm AFM}$ in the normal state is plotted against $x$.\cite{KobayashiPRB2013} 
In order to shed light on an evolution of electronic state as $x$ increases, we have examined the $x$ dependence of $N(E_{\rm F})$, which can be evaluated from $K_s$ in the normal state using both the relations of $K_s(x)=K(x)-K_{\rm chem}(=0.04\%)$ and $K_s\propto ^{31}\!$$A_{\rm hf}(q$=$0)N(E_{\rm F})$.
As shown in Fig. \ref{PhaseDiagram}(e), $N(E_{\rm F})$ gradually decreases with increasing $x$ due to a possible increase of the bandwidth with $x$ in Sr122AsP, since the P substitution shortens both  the Fe-Fe and Fe-$Pn$ bonding lengths. 

Next we compare this P-substitution change in $N(E_{\rm F})$ with those of Ba122AsP and La1111AsP as in Fig. \ref{PhaseDiagram}(b).
Note that the $K_s\propto^{31}\!$$A_{\rm hf}N(E_{\rm F})$ for Sr122AsP is roughly 1.4 times larger than that of Ba122AsP over the whole samples. 
When assuming that a ratio ($^{31}\!A_{\rm hf}^{\rm Sr}$/$^{31}\!A_{\rm hf}^{\rm Ba}$) is the same as $(^{75}\!A_{\rm hf}^{\rm Sr}$/$^{75}\!A_{\rm hf}^{\rm Ba})\sim$ 1.1,\cite{Kitagawa1,KitagawaSr122} the ratio $N(E_{\rm F})^{\rm Sr}/N(E_{\rm F})^{\rm Ba}$ is estimated to be $\sim$1.3, which is in accord with a value estimated from the band calculation for $M$Fe$_2$As$_2$($M$=Ba,Sr).\cite{Krellner} 
This is considered because the Fermi surface along the $k_z$ direction is more significantly warped for Sr122AsP than for Ba122AsP, that induces the larger DOS at $E_{\rm F}$.\cite{SuzukiARPES}
By contrast, the $N(E_{\rm F})$ in La1111AsP increases markedly for $x~>$~0.6,\cite{Miyake,MukudaJPSJ2014} since the Fermi surface originating from 3$d_{z^2}$ orbital gives rise to a large peak of $N(E_{\rm F})$ for $x~>$~0.6.  

Although $K(T)$ for all the samples stays constant in the normal state, $(1/T_1T)$s for 0.1 $\le$ $x$ $\le$ 0.5 are markedly enhanced due to the development of low-energy AFM-SFs upon cooling as seen in Fig.\ref{1/T1TvsT}(b). According to the previous studies on Fe-pnictides,\cite{Ning,NakaiPRB,Dulguun,KinouchiPRB,MukudaPRB2014,Imai,Oka,Hirano} we assume two-dimensional (2D) AFM-SFs model that gives 
the relation of $(1/T_1T)_{\rm AFM}\propto \chi_{\rm Q}(T) \propto 1/(T+\theta)$ if a system is close to AFM QCP at which AFM order collapses.\cite{Moriya} 
Here, the staggered susceptibility $\chi_{\rm Q}(T)$ with a wave vector ${\bm q}$=${\bm Q_{\rm AF}}$ follows the Curie-Weiss law. 
The $\theta$ is a measure of how close a system is to the AFM QCP and hence $(1/T_1T)_{\rm AFM}$ at $\theta=0$ diverges towards $T \rightarrow 0$.
As shown by the solid curves in Fig.~\ref{1/T1TvsT}(b), the $T$ dependence of $1/T_{1}T$ for $0 \le x\le 0.5$ can be reproduced by the relation of $1/T_{1}T= a/(T+\theta)+b$. Here, $(1/T_1T)_{\rm AFM}\equiv~a/(T+\theta)$ and $(1/T_1T)_{0}\equiv b$. 
Figure \ref{PhaseDiagram}(a) shows the contour plot of $(1/T_1T)_{\rm AFM}$ against $x$, which presents how the  AFM-SFs develop in the $x$-$T$ plane. The $T_N$ in the figure is determined by the rapid decrease of the NMR intensity, which was slightly lower than the structural transition at $T_s$ determined by the anomaly in resistivity.\cite{KobayashiPRB2013} 
The fitting parameters $\theta$, $a$, and $b$ for each $x$ are summarized in Figs. \ref{PhaseDiagram}(b), \ref{PhaseDiagram}(c),  and \ref{PhaseDiagram}(d), respectively.
Most remarkably, at $x\sim$0.35  $\theta$ approaches to zero, indicating that the AFM-SFs are critically enhanced towards $T \rightarrow 0$. 
Note that  the  $\theta$ markedly increases in going from $x$=0.35 to 0.6 as AFM-SFs become weak,  and no trace of AFM-SFs was seen at $x\ge$0.6, where the Curie-Weiss term $(1/T_1T)_{\rm AFM}$ is not resolved. 
Hence it is noteworthy that the onset and increase of $T_c$ are apparently associated with the emergence and enhancement of AFM-SFs, respectively. 

The parameter $a$ is a measure of the spectral weight of AFM-SFs at ${\bm q}$=${\bm Q_{\rm AF}}$ at low energies, that is, $\chi(Q,\omega_0)/\omega_0|_{\omega_0\rightarrow 0}$. 
When $a^*$(Ba) was evaluated from $a$(Ba)\cite{NakaiPRL} using the ratio of $(A_{\rm hf}^{\rm Sr}/A_{\rm hf}^{\rm Ba})^2$,  the $a$(Sr) and $a^*$(Ba) at $x$=0.35 for both the compounds are estimated to be  comparable. 
On the other hand, as shown in Fig. \ref{PhaseDiagram}(d), the value of $b$(Sr) is much larger than $b^*$(Ba). 
When noting that $(1/T_1T)_{0}\propto K_s^2$, the plot of $\sqrt{b}$ vs $K_s$ reveals a linear relation for Sr122AsP and Ba122AsP as indicated in Fig. \ref{PhaseDiagram}(f), demonstrating 
that the $b$ in the relation $1/T_{1}T= a/(T+\theta)+b$ is in proportion to $N(E_{\rm F})^2$ in fact.

\subsection{Evidence for unconventional SC properties} 

\begin{figure}[t]
\centering
\includegraphics[width=7cm]{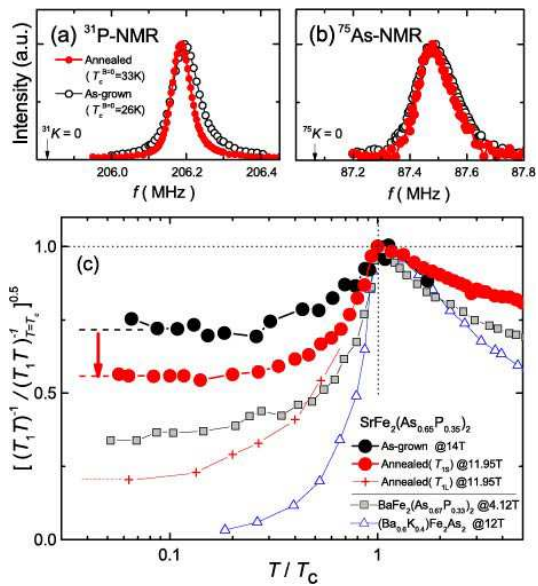}
\caption[]{(Color online) (a) $^{31}\!$P- and (b) $^{75}\!$As-NMR spectra for the annealed and as-grown samples of $x$=0.35 at $T$=70 K and $B_0\sim$11.95 T perpendicular to the $c$-axis.  
(c) Plots of $\sqrt{(1/T_1T)/(1/T_1T)_{T=T_c}}=N_{\rm res}/N_0$ versus $T/T_c$ for the as-grown sample (black circle) and the annealed one (red circle and cross), along with the results for Ba122AsP (square  mark)\cite{NakaiPRB} and BaK122(triangle mark).\cite{Yashima} 
} 
\label{anneal_NMR}
\end{figure}

The $T_c$ at $x$=0.35 is significantly enhanced from 26 K up to 33 K by annealing the as-grown sample.\cite{KobayashiPRB2013} 
As indicated in Fig. \ref{anneal_NMR}(a), the $^{31}\!$P-NMR spectra are narrower for the annealed one than for the as-grown one due to the better homogeneity for the former. 
This is also the case in the comparison with the $^{75}\!$As ($I$=3/2)-NMR spectra in Fig. \ref{anneal_NMR}(b). 
These NMR spectra may indicate that  the homogeneity of the electronic states and/or local structure of Fe$Pn$ tetrahedron are improved through the annealing process of the sample. 

Next we deal with the annealing effect on the SC properties of Sr122AsP at $x$=0.35. 
Figure~\ref{anneal_NMR}(c) shows plots of $^{31}$P-NMR $1/T_1T$ versus $T/T_c$ for Sr122AsP along with the result for Ba122AsP at $x$=0.33 (square mark) \cite{NakaiPRB} with $T_c$=30 K and Ba$_{0.6}$K$_{0.4}$Fe$_2$As$_2$(BaK122)\cite{Yashima} (triangle mark) with $T_c$=38 K. 
The $1/T_1T$s for Sr122AsP decrease steeply below $T_c$ without the coherence peak and follow a constant behavior below $T/T_c<$0.5 at respective fields $B_0=14$ T and 12 T for the as-grown (black circle)\cite{Dulguun} and annealed (red circle and cross mark) samples. 
The values of $1/T_1T$ normalized by the value at $T_c$ remain a finite. 
The previous work on the as-grown sample revealed that although it shows a weak $B_0$ dependence, it still remains a finite even in a low $B_0$ limit at $T\sim0.2 T_c$, pointing to a gapless SC state dominated by a large contribution of low-lying quasiparticle excitations at $E_{\rm F}$.\cite{Dulguun} 
By contrast, the $1/T_1T$ in BaK122  follows the power-law $T$ dependence without any constant behavior even under the high external field (12 T), which was consistently accounted for by the fully-gapped $s_{\pm}$-wave model.\cite{Yashima}
As for the case of the gapless SC state, the $1/T_1T$ is related to the square of RDOS at $E_{\rm F}$ $N_{\rm res}(E_{\rm F})^2$, and hence the fraction of RDOS ($N_{\rm res}/N_0$) to a normal-state DOS $N_0$ is given by $\sqrt{(1/T_1T)/(1/T_1T)_{T=T_c}}$.
Using this relation, $N_{\rm res}/N_0$=0.56, 0.66 and 0.72 were evaluated in the previous work for the as-grown one at respective magnetic fields $B_0$=1, 2.9 and 14 T\cite{Dulguun}, which was corroborated by the specific-heat measurements, deducing a comparable magnitude of RDOS at the same field. 
These values are much larger than $N_{\rm res}/N_0\sim$0.34 at $B_0\sim$~4 T for the nodal-gap state in Ba122AsP\cite{NakaiPRB} and $N_{\rm res}/N_0$=0 for the fully gapped state in BaK122.\cite{Yashima} 

As seen in Fig. \ref{anneal_NMR}(c), the $1/T_1T$ for the annealed one decreases steeply without the coherence peak below $T_c(B_0)\sim$30 K under $B_0=11.95$ T, followed by the constant behavior below $T/T_c<$ 0.5 as well as for the as-grown one. 
Note here that  a single component of $T_1$ in the normal state is deduced from the nuclear relaxation curve $m(t) = [M_0-M(t)]/M_0 = \exp(-t/T_1)$ for $^{31}\!$P($I=1/2$).
On the other hand, the nuclear recovery curve with a multicomponent of $T_1$ is assumed in the SC state well below $T_c$ as $m(t) = [M_0-M(t)]/M_0 =\sum_i~A_i\exp(-t/T_{1i})$. 
This distribution in $T_1$ may be attributed to the spatial distribution of electronic states in the vortex state. Here we assume tentatively a two-component model with $T_{1\rm S}$ and $T_{1\rm L}$.
The respective values of $N^{\rm S}_{\rm res}/N_0$ and $N^{\rm L}_{\rm res}/N_0$ are estimated to be 0.56 and 0.2 from $1/T_{1\rm S}T$ and $1/T_{1\rm L}T$. 
Namely, the result demonstrates that the respective fractions of RDOS at $^{31}\!$P site around and far from vortex cores are spatially distributed from 0.56 to 0.20. The RDOS may be largely induced in the vicinity of the vortex cores where the order parameter is remarkably depressed. 
We note that $N^{\rm L}_{\rm res}/N_0\sim$0.2 far from the vortex cores is roughly consistent with $N^{\gamma_e}_{\rm res}/N_0$=0.18 estimated from the residual electronic specific heat $N^{\gamma_e}_{\rm res}$ at low $T$ limit at zero field without the vortices.\cite{KobayashiPRB2013} 

As a result, the increase of $T_c$ at $x$=0.35 from $T_c$=26 K  to 33 K is attributed to the reduction of  the lattice defects by annealing the as-grown mono-crystalline sample. 
This is inconsistent with the case of an isotropic conventional $s$-wave SCs, where nonmagnetic scatterers like lattice defects does not suppress $T_c$, which is known as Anderson's theorem.
The RDOS at $E_{\rm F}$ in the SC state of Sr122AsP becomes much smaller for the annealed mono-crystalline sample than for the as-grown one, suggesting that some inhomogeneity and/or imperfection increase the RDOS within some of the multiple SC gaps with nodes. 
This is consistent with the case expected for unconventional $s_\pm$ wave SC state with nodal gap\cite{Hirschfeld}.

\section{Discussion} 


\begin{figure}[htbp]
\centering
\includegraphics[width=6cm]{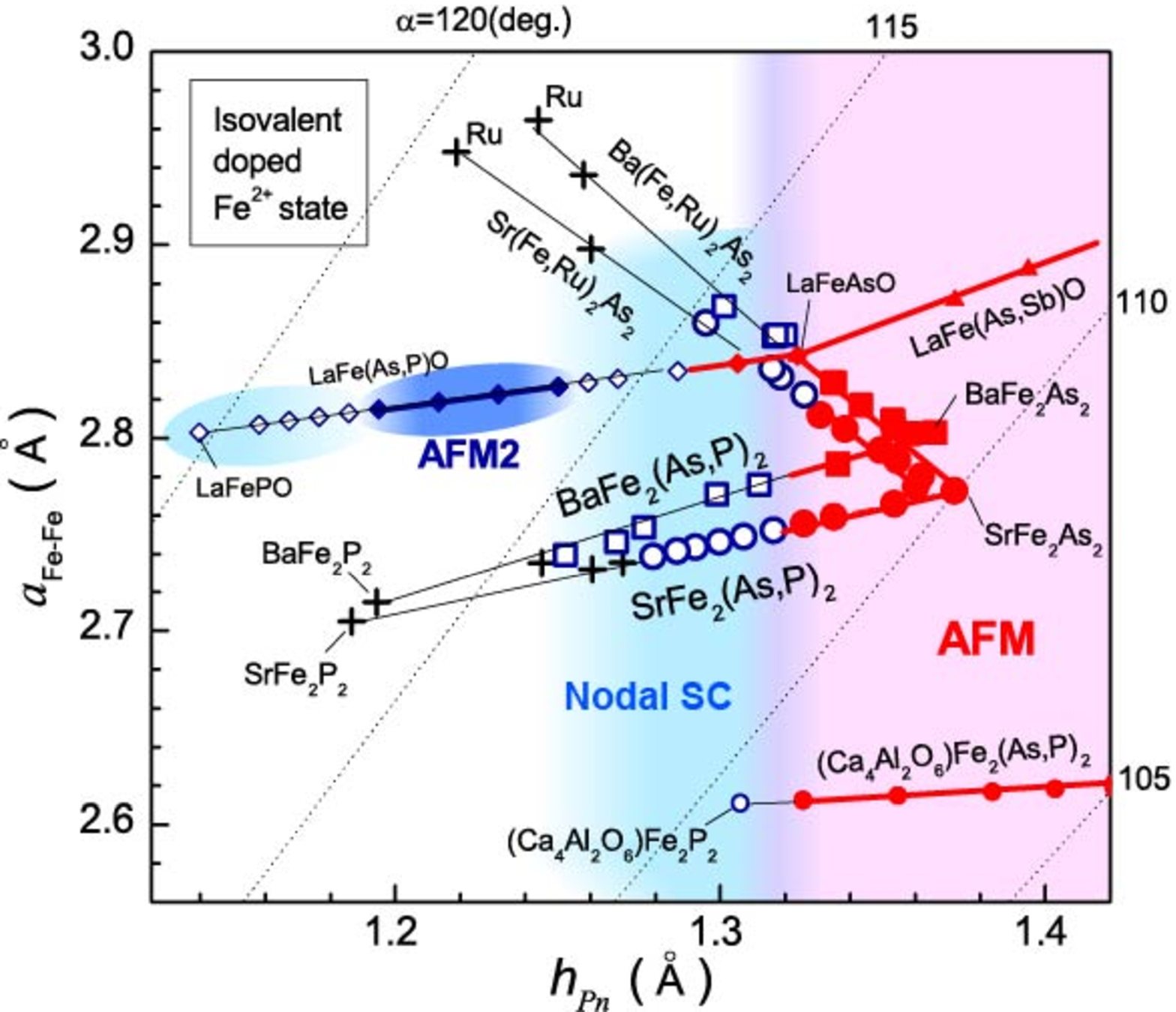}
\includegraphics[width=6cm]{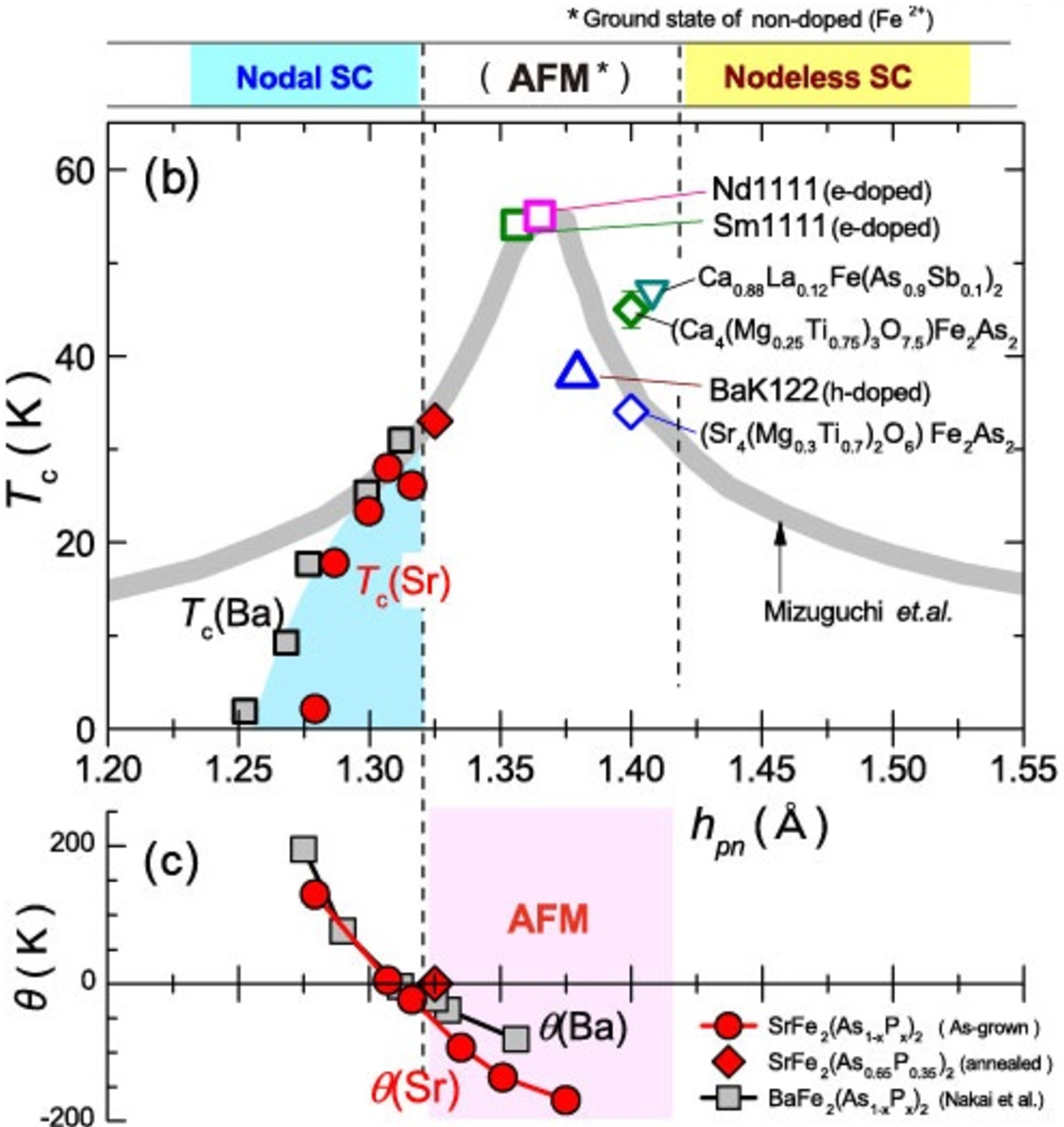}
\includegraphics[width=6cm]{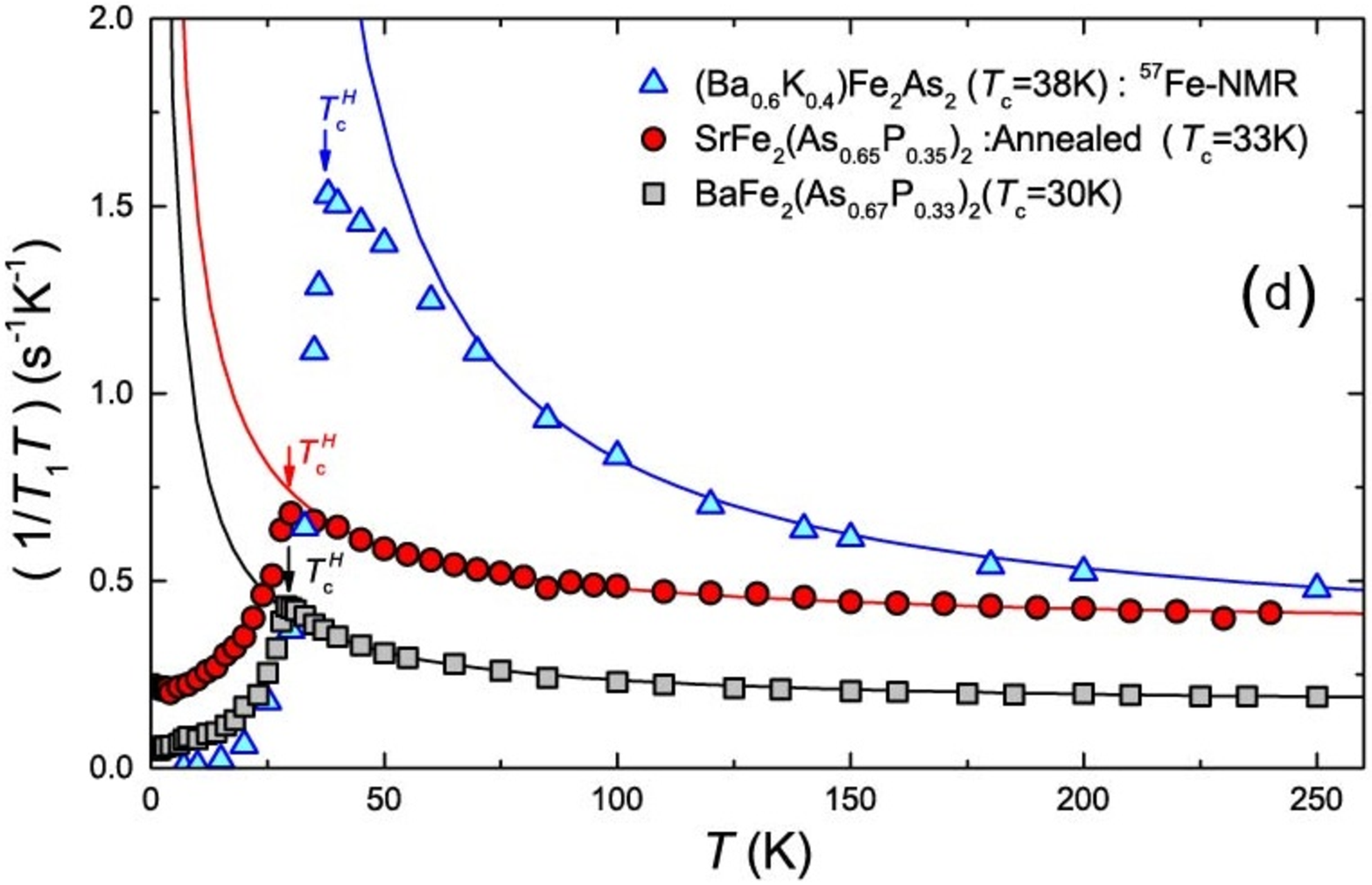}
\caption[]{(Color online)  
(a) Ground states of isovalent substituted Fe-pnictides plotted as functions of  $h_{Pn}$ and the inplane Fe-Fe distance $a_{\rm Fe-Fe}$. 
The filled and empty symbols denote the AFM ordered phase and the SC phase, respectively, for  SrFe(As,P)$_2$\cite{KobayashiPRB2013}, Sr(Fe,Ru)$_2$As$_2$\cite{Schnelle}, BaFe(As,P)$_2$\cite{Kasahara}, Ba(Fe,Ru)$_2$As$_2$\cite{Rullier-Albenque,Qiu,Nath}, 
LaFe(As,P/Sb)O\cite{C.Wang,LaiPRB,MukudaJPSJ2014,Carlsson}, and  Ca$_4$Al$_2$O$_6$Fe$_2$(As,P)$_2$\cite{Shirage,KinouchiPRB}. 
Plots of (b) $T_c$ and (c) $\theta$ versus pnictogen height $h_{Pn}$ for Sr122AsP and Ba122AsP. 
Here the broad line is the $h_{Pn}$ dependence of $T_c$ with respect to various {\it doped} Fe-pnictides reported by Mizuguchi {\it et al.}\cite{Mizuguchi}  
Empty symbols are plots of $T_c$s of some {\it doped} Fe-pnictides revealing $T_c$ higher than the maximum of $T_c$ at QCP around $h_{Pn}$~$\sim$~1.32\AA:  
These samples are the electron-doped SmFeAs(O,F) ($T_c\sim$55 K),\cite{Ren1}  the Ca$_4$(Mg$_{0.25}$Ti$_{0.75}$)$_3$O$_{7.5}$Fe$_2$As$_2$ ($T_c\sim$45 K),\cite{Ogino_43822} the Sr$_4$(Mg$_{0.3}$Ti$_{0.7}$)$_2$O$_{6}$Fe$_2$As$_2$ ($T_c\sim$34 K),\cite{Sato}
the Ca$_{0.88}$La$_{0.12}$Fe(As$_{0.9}$Sb$_{0.1}$)$_2$ ($T_c\sim$47 K),\cite{Kudo_47K} and the hole doped BaK122 ($T_c\sim$38 K).\cite{Rotter}
(d) $T$ dependence of $^{31}\!$P-NMR $(T_1T)^{-1}$ in the normal state for Sr122AsP with $T_c$=33 K and  Ba122AsP with $T_c$=30 K,\cite{NakaiPRL} along with $^{57}\!$Fe-NMR $(T_1T)^{-1}$ of hole-doped BaK122 with $T_c$=38 K that is the highest $T_c$ among 122 compounds.\cite{Yashima} 
The data of BaK122 with optimum $h_{Pn}$ deviate below $\sim$70K from the solid curve that assumes the simple Curie-Weiss law $1/T_1T=a/(T+\theta)+b$. 
}
\label{h_pn}
\end{figure}

The present NMR study  has unraveled that $T_c$ reaches the highest value of 33 K at $x$=0.35 in the presence of the enhanced AFM-SFs around  QCP, and no SC transition takes place for the sample at $x>$0.6 without any trace of AFM-SFs. 
Consequently, the onset and increase of $T_c$ are apparently associated with the emergence and enhancement of AFM-SFs in the vicinity of AFM order, respectively. 
In this section, we discuss the relationship between the electronic states and the local  lattice parameters of Fe-$Pn$ layer. 

Figure \ref{h_pn}(a) indicates the AFM/SC ground states plotted as functions of  $h_{Pn}$ and the inplane Fe-Fe distance ($a_{\rm Fe-Fe}$) for  Sr122AsP, along with the data on the various isovalent-substitutied Fe-pnictides\cite{KinouchiPRB}, which are characterized by Fe$^{2+}$ state. 
The phase boundary between the AFM and the nodal SC phases appears at $h_{Pn}\sim$1.32\AA\  for Sr122AsP. 
The similar phase transition occurs at the same $h_{Pn}$  in the isovalent Ru-substituted series  Sr(Fe,Ru)$_2$As$_2$\cite{Schnelle}, even though the substituted atomic site and  $a_{\rm Fe-Fe}$ differ significantly from those of Sr122AsP. 
This trend is widely seen not only in BaFe(As,P)$_2$\cite{Kasahara} and  Ba(Fe,Ru)$_2$As$_2$\cite{Rullier-Albenque,Qiu,Nath}, but also in  LaFe(As,P/Sb)O\cite{C.Wang,LaiPRB,MukudaJPSJ2014,Carlsson}, Ca$_4$Al$_2$O$_6$Fe$_2$(As,P)$_2$\cite{Shirage,KinouchiPRB}, as shown in Fig.  \ref{h_pn}(a). 
Figures \ref{h_pn}(b) and \ref{h_pn}(c) present the $T_c$ and $\theta$ versus $h_{Pn}$ for Sr(Ba)122AsP, respectively. 
It is noteworthy that the $\theta\sim 0$ appears around $h_{Pn}\sim$1.32\AA~both for Sr122AsP and Ba122AsP, although the $T_N$ of  the parent compound SrFe$_2$As$_2$ is much higher than that of BaFe$_2$As$_2$. 
These results suggest that the pnictogen height  is a dominant factor to determine the phase boundary between AFM order and nodal SC phases in the isovalent substitution compounds\cite{KinouchiPRB}.  

Next we address on additional key-factor to raise $T_c$ further in Fe-based pnictides, including the extra  {\it electron- or hole-doped} compounds.
The role of AFM-SFs for enhancing $T_c$ has been widely recognized in various Fe-pnictides such as Ba(Fe,Co/Ni)$_2$As$_2$,\cite{Ning,Zhou}, Fe(Se,Te),\cite{Imai,YShimizu} BaFe$_2$(As,P)$_2$,\cite{NakaiPRL}, Na(Fe,Co)As,\cite{Ji} LaFe(As,P)(O,F),\cite{Oka,SKitagawa2014,MukudaPRB2014} and so on. 
The highest $T_c$ state of Fe-pnictides emerges at $T_c\sim$ 55 K for the electron-doped $R$1111OF\cite{Ren1},  as the result of the depression of the AFM order by the extra electron doping  through the substitution of F$^{1-}$ for O$^{2-}$. 
As plotted in Fig. \ref{h_pn}(b),  some samples such as$R$1111OF,  electron-doped perovskite-block-type compound Ca$_4$(Mg$_{0.25}$Ti$_{0.75}$)$_3$O$_{7.5}$Fe$_2$As$_2$ ($T_c$$\sim$ 45 K),\cite{Ogino_43822} electron-doped 112-type compound Ca$_{0.88}$La$_{0.12}$Fe(As$_{0.9}$Sb$_{0.1}$)$_2$ ($T_c\sim$47K),\cite{Kudo_47K} and the hole doped 122 compound BaK122($T_c$=38K)\cite{Rotter} are always characterized by $h_{Pn}=1.35$~$\sim$~1.4\AA~and the bond angle of 
$Pn$-Fe-$Pn$~$\alpha$~$\sim$~109.5$^\circ$, in which the local lattice parameters of the FeAs layer are close to the values for regular FeAs$_4$ tetrahedron\cite{Mizuguchi,C.H.Lee}.
Note that in such compounds the $T$ dependence of $1/T_1T$ tends to saturate towards $T_c$  and/or show the broad maximum above $T_c$, as reported in  (Y$_{0.95}$La$_{0.05}$)1111($T_c$=50 K),\cite{MukudaPRL,MukudaPRB2014} Ca$_4$(Mg,Ti)$_3$Fe$_2$As$_2$O$_{8-y}$ ($T_c$= 45 K),\cite{Tomita}  Sr$_4$(Mg$_{0.3}$Ti$_{0.7}$)$_2$O$_6$Fe$_2$As$_2$ ($T_c$=34 K)\cite{Yamamoto}, and so on.
As for the 122 series,  although the low energy AFM-SFs are significantly observed in BaK122\cite{Yashima,Li,Hirano,Matano},  the relationship between the $T_c$ and QCP in the phase diagram was not clearly revealed in  Ba$_{1-x}$K$_{x}$Fe$_2$As$_2$\cite{Hirano}. 
In fact, as shown in Fig.\ref{h_pn}(d), for example,  the $T$ dependence of $1/T_1T$ for $x\ge$0.4 does not simply follow the  Curie-Weiss behavior of $1/T_1T=a/(T+\theta)+b$\cite{Yashima,Hirano}, in contrast with the cases of Sr122AsP and Ba122AsP\cite{NakaiPRB}. 
Since the local structure of BaK122 is optimized with the regular FeAs$_4$ tetrahedron, we infer that the high degeneracy of $3d$ orbitals may give some influences on the evolution of the low energy AFM-SFs. 

In this context, we remark that the AFM-SFs play an important role for the onset of superconductivity in Fe-pnictides widely, but they are not an unique factor for reaching $T_c$=55 K in these compounds. 
When noting that the highest value of $T_c$ in Fe-based pnictides occurs when the local structure is optimized with the regular FeAs$_4$ tetrahedron, some multiorbital-derived SFs  may play an important role due to the high degeneracy of $3d$ orbitals. 
Recently, one scenario is theoretically proposed that the high-energy multiorbital-derived SFs due to the orbital degeneracy is a mediator for the sign reversal $s_\pm$-wave superconductor in high $T_c$ Fe-based pnictides.\cite{Arai}
The other possible scenario is that the cooperation of ferro-orbital fluctuation at $q\sim0$ and AFM-SFs at $q=Q_{\rm AF}$ enhances the $T_c$ further within the framework of $s_\pm$-wave Cooper pairing state\cite{Ono}. 
Some multi-orbital fluctuations relevant with the degeneracy of $3d$ orbitals may play some role for the onset of superconductivity\cite{Onari,Saito}, since the spin and orbital degrees of freedom might be coupled one another.

\section{Summary} 

Systematic $^{31}\!$P-NMR  studies on the single crystallines SrFe$_2$(As$_{1-x}$P$_{x}$)$_2$ have revealed that the normal-state property is dominated by the development of AFM-SFs for the compound at $x$=0.35, but not for the compound at $x$=0.6 that exhibits no SC transition.  We remark that the onset and increase of $T_c$ are apparently associated with the emergence and enhancement of AFM-SFs, respectively. 

The $T_c$ for the sample at $x$=0.35 is enhanced from 26 K up to 33 K by annealing the as-grown sample, resulting in the significant reduction of RDOS. The fact that the RDOS remains finite even for the annealed one indicates that the subtle impurity scattering may induce the RDOS at $E_{\rm F}$ even in the very clean sample of Sr122AsP. These normal and SC properties are consistently accounted for by the $s_\pm$-wave Cooper pairing model mediated by the AFM SFs.\cite{Mazin,Kuroki1} 

We have also discussed about other key-ingredients besides the AFM-SFs to increase  $T_c$ further. 
The highest value of $T_c$=55 K in $R$1111OF is higher than the maximum of $T_c$$\sim$33 K at QCP around $h_{Pn}$~$\sim$~1.32\AA. It is known that the local structure of these compounds is optimized as the regular FeAs$_4$ tetrahedron. Thus, the optimization of the local structural parameters with the regular Fe$Pn$ tetrahedron are necessary for bringing about the highest $T_c$ in $R$1111OF through the additional effect such as the spin-oribital coupled fluctuations over ranging low to high energies\cite{Arai} and/or some multiorbital fluctuations due to the large degeneracy of $3d$ orbitals. 

\section*{Acknowledgements}

{\footnotesize 
We thank K. Kuroki, H. Usui, K. Suzuki, T. Yoshida, A. Fujimori, and K. Ishida for valuable comments. 
This work was supported by Grants-in-Aid for Scientific Research (Nos. 26400356 and  26610102) from the Ministry of Education, Culture, Sports, Science and Technology (MEXT), Japan.
}

\end{document}